\documentclass[9pt,twocolumn,twoside]{opticajnl}
\journal{opticajournal} 

\setboolean{shortarticle}{false}

\usepackage{lineno}

\title{Cryogenic Focus Measurement System for a Wide-Field Infrared Space Telescope}

\author[1,*]{Samuel S. Condon}
\author[1]{Stephen Padin}
\author[1]{James Bock}
\author[1]{Howard Hui}
\author[1]{Phillip Korngut}
\author[1]{Chi Nguyen}
\author[1]{Jordan Otsby}

\affil[1]{California Institute of Technology, Cahill Center for Astronomy and Astrophysics, 1216 East California Boulevard, Pasadena, CA, 91125, USA}
\affil[*]{scondon@caltech.edu}

\begin{abstract}

    We describe a technique for measuring focus errors in a cryogenic, wide-field, near-infrared space telescope. The measurements are made with a collimator looking through a large vacuum window, with a reflective cold filter to reduce background thermal infrared loading on the detectors and optics. The vacuum window and cold filter introduce wavefront error which we characterize using an autocollimating microscope. For the $200\textrm{ mm}$ diameter aperture $f/3$ space telescope, SPHEREx, we achieve a focus position measurement with $\sim15 \textrm{ }\mu \textrm{m systematic}$ and $\sim 5\textrm{ }\mu \textrm{m statistical}$ error.
  
\end{abstract}

\setboolean{displaycopyright}{false} 

\begin{document}

\maketitle

\section{Introduction}

    Effective observation of low brightness astrophysical sources in the infrared requires low detector dark current and instrumental thermal emission necessitating the use of cryogenic cooling systems in infrared telescopes. Additionally, telescopes are often equipped with a mechanism to adjust the focus, based on in situ measurements of the image size. In a cryogenic space telescope, a focus mechanism presents a single point mechanical failure such that it can be advantageous to fly a fixed focus system. In this case, the focus must be accurately measured and set on the ground before the telescope is launched. To account for mechanical deformation due to cooling and its effect on the focus, a ground-based focus measurement system must cool the telescope to cryogenic operating temperatures. This paper describes a laboratory focus measurement system for SPHEREx, a small, fixed-focus, wide-field, near-infrared cryogenic space telescope \cite{SPHEREx}, which is passively cooled to 60 K during operation in a low-earth orbit.

    Our focus measurement system cools the telescope in a cryogenic test chamber and couples light from an external collimator through a vacuum window. A cold filter reduces infrared loading from the laboratory $300 \textrm{ K}$ background, which would otherwise saturate the detectors. A reflective coating on the inside of the vacuum window reduces the cold load from the 80 K filter. Difficulties in this focus measurement arise from (1) measuring the wavefront error introduced by the cold filter because its optical transmission is low and (2) the wide field of view of the telescope, which results in a very large vacuum window and filter.
 
    The following sections describe the SPHEREx telescope, the setup for focus measurements in the laboratory, design details for the test chamber vacuum window and cold filter, calibration of the collimator and coupling optics, sources of measurement error, results from SPHEREx focus measurements, and an additional “warm” focus measurement technique.
    
\section{SPHEREx}

    The Spectro-Photometer for the History of the Universe, Epoch of Reionization and Ices Explorer (SPHEREx) is a NASA medium-class Explorers satellite that will survey the entire sky at near-infrared wavelengths with a $6.2"$ angular resolution and a spectral resolution of $\textrm{R}=\frac{\lambda}{\Delta \lambda} = 35-130$. The telescope is an $f/3$ anastigmat with three free-form off-axis mirrors, a 200 mm diameter aperture, and a $11.3^{\circ}\times3.5^{\circ}$ field of view. The pupil is located at the secondary mirror, which results in separated beam patterns over the primary and tertiary mirrors. The housing and mirrors are aluminum and a dichroic beamsplitter feeds two focal planes covering $\lambda=0.75\textrm{--}2.42\textrm{ }\mu\textrm{m}$ (reflected channel) and $\lambda=2.42\textrm{--}5\textrm{ }\mu\textrm{m}$ (transmitted channel) \cite{Frater}. Each focal plane contains three Teledyne H2RG detector arrays with $2048\times2048$ pixels \cite{Blank_H2RG}. A linear variable filter mounted immediately in front of each detector sets the observing wavelength and spectral resolution \cite{LVF}. Table \ref{tab:InstrumentParameters} gives a summary of the instrument parameters.
    
		\begin{table}[h]
			\centering
			\caption{SPHEREx instrument parameters. \label{tab:InstrumentParameters}}	
			\begin{tabular}{p{0.3\columnwidth}p{0.6\columnwidth}}
				\hline
				Parameter & Value
				\tabularnewline
				\hline
				Aperture & $200\textrm{ mm}$
				\tabularnewline
				Focal length & $600\textrm{ mm}$
				\tabularnewline
				Field of view & $11.3^{\circ}\times3.5^{\circ}$
				\tabularnewline
				\everymath{\displaystyle}
				Detectors & $6\!\!\!\times\!\!\!\textrm{H2RG:  } 2048\!\!\!\times\!\!\!2048\textrm{, }18\!\!\!\times\!\!\!18\textrm{ }\mu\textrm{m pixels} (6.2"\times6.2")$
				\tabularnewline
				\everymath{\displaystyle}
				Wavelength range & Band 1: $\lambda=0.75\textrm{--}1.11\textrm{ }\mu\textrm{m}$, $R=41$
				\tabularnewline
				\everymath{\displaystyle}
				& Band 2: $\lambda=1.11\textrm{--}1.64\textrm{ }\mu\textrm{m}$, $R=41$
				\tabularnewline
				\everymath{\displaystyle}
				& Band 3: $\lambda=1.64\textrm{--}2.42\textrm{ }\mu\textrm{m}$, $R=41$
				\tabularnewline
				\everymath{\displaystyle}
				& Band 4: $\lambda=2.42\textrm{--}3.82\textrm{ }\mu\textrm{m}$, $R=35$
				\tabularnewline
				\everymath{\displaystyle}
				& Band 5: $\lambda=3.82\textrm{--}4.42\textrm{ }\mu\textrm{m}$, $R=110$
				\tabularnewline
				\everymath{\displaystyle}
				& Band 6: $\lambda=4.42\textrm{--}5.00\textrm{ }\mu\textrm{m}$, $R=130$
				\tabularnewline
				\hline
			\end{tabular}
		\end{table}

    The three free-form off-axis mirrors are aligned with wavefront measurements using an interferometer and an array of spherical reflectors in place of the detectors while the focal planes are aligned using measurements with the collimator configuration described in this paper. Alignment of each focal plane must be set such that the point spread function remains smaller than a detector pixel. In the short wavelength bands, this means that the diameter of the geometric spot size, given by the focus error over the telescope focal ratio $(d = \delta / F = \delta / 3)$, must be less than the $18 \textrm{ } \mu \textrm{m}$ width of a SPHEREx detector pixel, placing a focus accuracy requirement of $< 50 \textrm{ } \mu \textrm{m}$. Due to diffraction, the focus accuracy may be $< 100 \textrm{ } \mu \textrm{m}$ in the long wavelength bands.

\section{Cold Focus Method and Test Setup} \label{sec:ColdFocusMeasurements}

\begin{figure}[h]
			\centering
			\includegraphics[scale=1]{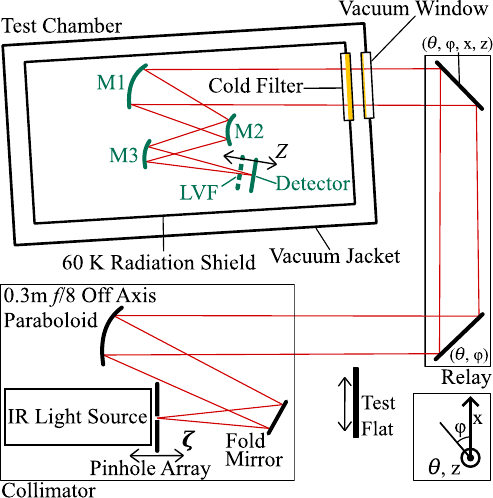}
			\caption{Collimator configuration for telescope focus measurements. Green lines indicate the SPHEREx optics. The position of the pinhole array at collimator focus is denoted as $\zeta$ and the position of its image at the telescope focus is denoted as $Z$. Two steerable relay mirrors adjust the input angle into the telescope to measure the PSF over the entire field of view. The relay mirrors have six total degrees of freedom over the coordinates $x, z, \theta, \phi$ where $x, z$ are orthogonal linear displacements, $\theta$ is a rotation angle measured relative to $z$, and $\phi$ corresponds to a rotation in the plane of the page. A test flat can be inserted at the output of the collimator to determine the quality of the collimator beam as well as the position of collimator focus, $\zeta_{\textrm{col}}$.}
			\label{fig:TelFocusSetup}
		\end{figure}

        We use the collimator configuration shown in Fig. \ref{fig:TelFocusSetup} to measure focus error in the focal plane across the entire field of view of the telescope. We then install aluminum shims in the focal plane to shift the position of the surface of the detectors to minimize the focus error. After shimming, we confirm that focus shifts as expected by repeating the measurement. This iterative process can be continued until the focus performance meets requirements. In practice, we perform three sets of focus measurements, adjusting the detector positions using shims between the first and second sets. The results of this process are discussed in Section \ref{sec:Results}.

        We obtain the telescope focus error with \cite{Zemcov2013Ciber}:
  
		\begin{equation*} \label{eqn:FocusError}
			dZ = (\zeta_{\textrm{tel}} - \zeta_{\textrm{col}}) \left ( \frac{f_{\textrm{tel}}}{f_{\textrm{col}}} \right )^2
		\end{equation*}

		\noindent
        where $f_\textrm{tel}$ and $f_\textrm{col}$ are the focal length of the telescope and collimator, $\zeta_\textrm{col}$ is the position of the point source at collimator focus to generate a collimated beam inside the test chamber, and $\zeta_\textrm{tel}$ is the position of the point source at collimator focus to minimize the spot area at telescope focus. We use $Z$ to denote positions at the telescope focus. The lateral magnification from the telescope focus to the collimator focus is $m=f_\textrm{col}/f_\textrm{tel}$, and the longitudinal magnification is $m^2 = \left( 2.54 / 0.6 \right)^2 \approx 18$. A large $f_\textrm{col}$ is generally desirable because it gives large longitudinal magnification, which relaxes the tolerance on pinhole position measurements but also results in a long collimator setup that is more susceptible to seeing and mechanical instability.
        
		Our collimator utilizes a Space Optics Research Labs, 0.3 m off-axis parabolic mirror with $f_{\textrm{col}} = 2.54 \textrm{ m}$. Before the paraboloid, a small fold mirror moves focus away from the collimated beam to provide space for the pinhole array and light source. The light source is a Thorlabs SLS202L stabilized tungsten lamp with a roughly 2000 K blackbody spectrum over the $\lambda = 0.45 - 5.5 \; \mu$m band. The source mounts directly behind the pinhole array. For focus measurements with SPHEREx, we use a pinhole array containing 5 holes that mounts on a 3-axis stage with an absolute linear encoder (Heidenhain AT 1218) which reads pinhole position. The pinhole array samples the telescope PSF over a range of pixelizations, given SPHEREx's large $6.2'' \times 6.2''$ detector pixels. We use array hole diameters between $15 \textrm{ and } 100 \textrm{ } \mu\textrm{m}$ to set the light level.
		
\begin{figure}[H]
            \centering
            \includegraphics[scale=1]{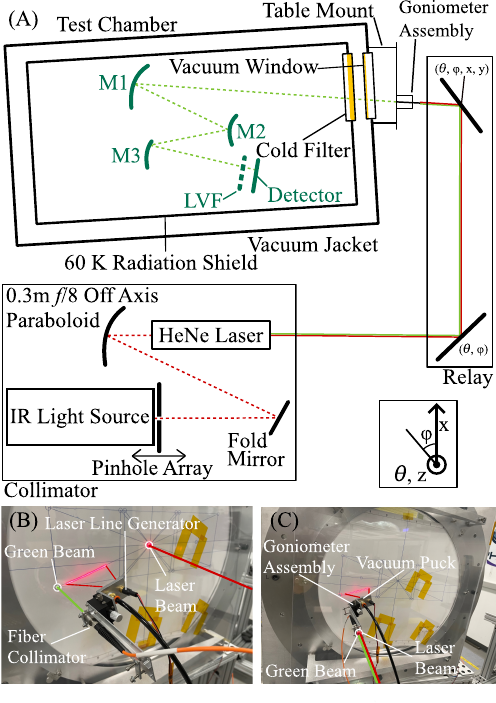}						
            \caption{Relay alignment setup to obtain axis tilts and offsets in our 3D model. (A) We align a red helium-neon (HeNe) laser with the collimator central axis and a green fiber-collimated LED with the telescope central ray for the desired field location. The green LED mounts on a "goniometer assembly" which consists of a mount to adjust the angle of the green LED beam, a vacuum puck to attach the LED to a plastic table mounted on the front of the vacuum window, and a laser line generator to aid in placing the LED beam coplanar with telescope boresight. By adjusting the relay such that the red laser and green beam are colinear, we place the collimator central axis in line with the telescope central ray. (B) Close view of the "goniometer assembly" mounted on the plastic table on the front of the vacuum window: Here we direct the green beam toward the telescope and adjust its angle and position to match a paper template which maps the expected interception of the central ray with the window. (C) After setting the position and angle of the goniometer assembly, we flip the green beam to trace the central ray through the relay flats.}
            \label{fig:RelayAlignment}
        \end{figure}

		A two-mirror relay couples the fixed collimator beam into the telescope entrance pupil. Uniform illumination of the entrance pupil is desired, which requires alignment of the collimator central axis with the telescope central ray for the desired field location. To achieve this for arbitrary field locations on the $11.3^{\circ} \times 3.5^{\circ}$ field of view, we developed a 3D solid model to trace rays from the collimator through the flats and to predict the telescope central ray projection. We obtain initial axis tilts and offsets in the model using the alignment setup of Figure \ref{fig:RelayAlignment}.		
  
\begin{figure}[H]
			\centering
			\includegraphics[scale=1]{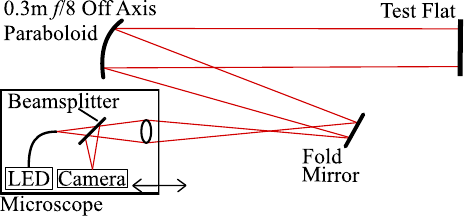}
			\caption{Autocollimating microscope setup to find collimator focus. We fiber couple an LED to the microscope, which re-images the tip of the fiber to create a point source at collimator focus. At the output of the off-axis-paraboloid, we place a test flat and inspect the returned image with a camera inside the microscope. The fiber and camera are the same distance from the microscope objective, so minimum spot area at the camera corresponds to the point source being at the collimator focus.}			
			\label{fig:ColFocusSetup}
		\end{figure}

		In order to measure the position of collimator focus, we use the autocollimating microscope setup shown in Figure \ref{fig:ColFocusSetup}. We take the minimum of a quadratic fit to spot area vs. microscope position as the collimator focus position. In this case, we calculate the area by simply counting the number of pixels above 10 or $25\%$ of the peak. The calculation runs fast, and it has good resolution because the spot area is always much larger than a 

        \noindent
        microscope camera pixel. Seeing, mechanical stability, and the choice of threshold for the spot area calculation affect the best-fit focus at the $5\textrm{ }\mu\textrm{m}$ level at telescope focus. From the size of the returned beam at best focus, the collimator provides a $5''$ PSF at a wavelength of $550 \textrm{ nm}$.

		After placing the microscope at collimator focus, we insert the pinhole array and adjust its position to minimize the area of the pinhole image in the microscope camera. This puts the pinhole array at collimator focus. We automate all of the microscope and pinhole operations using the SPHERExLabTools data acquisition and instrument control system \cite{SPHERExLabTools}. This makes it easy to check 
  
        \noindent
        collimator performance at regular intervals during a series of telescope focus measurements.

\begin{figure}[H]
        \centering
        \includegraphics[scale=1]{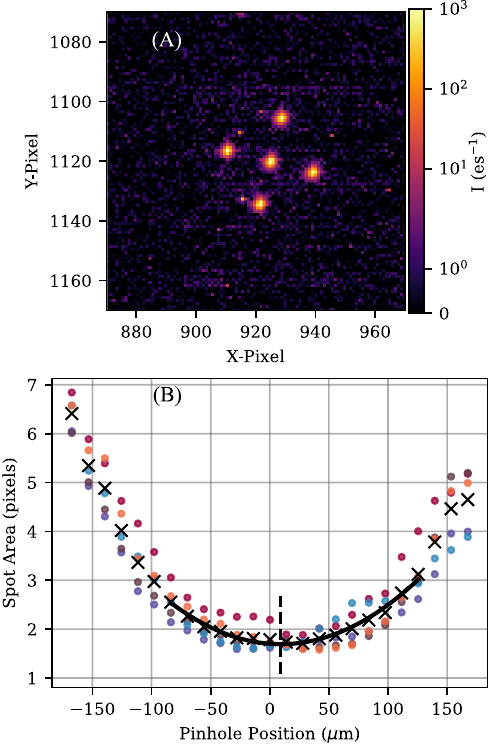}
        \caption{(A) Typical image of a five-spot pinhole array observed on the telescope focal plane. The largest spots in the center of the frame correspond to the spots from the pinhole array. Other features are from stray light or detector hot pixels.(B) Spot area versus pinhole position on telescope boresight at $\lambda = 1.35 \textrm{ } \mu \textrm{m}$. Colored dots correspond to using the area for each individual spot and black crosses indicate the average area for all five spots. The solid black line is a quadratic fit to a limited region of pinhole positions, centered about the minimum measured spot area and the dashed black line is the minimum of the fit. The horizontal axis corresponds to the displacement of the pinhole array measured in units at telescope focus.}
        \label{fig:TelFocusMeasurement}
    \end{figure}

		Figure ~\ref{fig:TelFocusMeasurement} shows a typical focus measurement where a fit to spot area vs. pinhole position determines the best focus position. To calculate the spot area, we use a Gaussian fit or encircled energy. These metrics work well with images that are coarsely sampled with large detector pixels. We fit a quadratic to the focus curves, using a subset of the measurements centered on best focus, or we find the center of the range of pinhole positions that corresponds to the number of pixels below some threshold value. Best focus positions from the different calculations typically agree within $\sim \!\! 6 \; \mu \textrm{m}$ at telescope focus. 
  
\section{Vacuum Window and Cold Filter} \label{sec:Window+Filter}

    The test chamber vacuum window is a $500\textrm{ mm}$ diameter clear aperture ($550\textrm{ mm}$ outside diameter), $20\textrm{ mm}$ thick, crystal sapphire plate. Sapphire has a large and flat transmission spectrum in the near-infrared ($80 \textrm{--} 85 \% \textrm{ from } \lambda = 0.75 \textrm{--} 5.0 \; \mu$m) and is mechanically very strong \cite{Sapphire}. The diameter is the largest readily available, but still vignettes $\sim \!\! 30\%$ of the telescope entrance pupil at the corners of the field of view. Sapphire of this size is only available in a-plane, which is birefringent. The birefringence, combined with the $3-5 '$ window wedge, results in $\sim \!\! 2''$ smearing of images generated with unpolarized light, which is smaller than a detector pixel. One must take care in mounting the plate since the crystal can easily chip or cleave. The vacuum window has an o-ring seal with a hard, square, rubber gasket just inside the o-ring to prevent the sapphire from touching the metal flange that supports the window. The window thickness was chosen to give a safety factor of 10 with vacuum loading.		

\begin{figure}[H]
        \centering
        \includegraphics[scale=1]{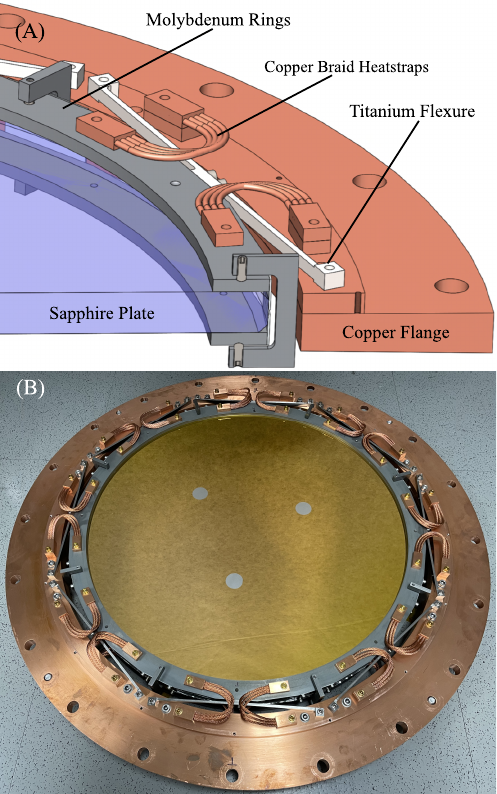}
        \caption{(A) Cold filter assembly cross section. (B) Cold filter assembly with the gold-coated side of the sapphire face up and the Hartmann mask spots uncovered.}
        \label{fig:ColdFilterAssembly}
    \end{figure}

\begin{figure}[H]
        \includegraphics[scale=1]{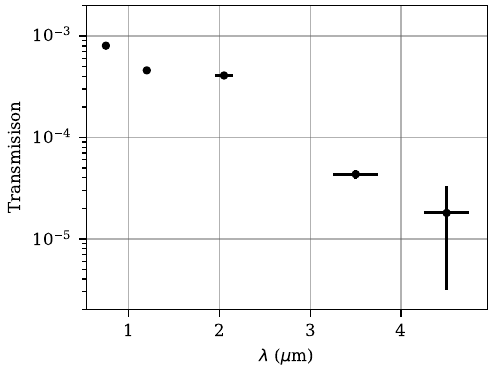}
        \caption{Measured transmission of the coated cold filter assembly. Horizontal errorbars indicate the FWHM bandwidth of the light source used at each point.}
        \label{fig:FilterTransmission}
    \end{figure}

\begin{figure}[h]
        \includegraphics[scale=1]{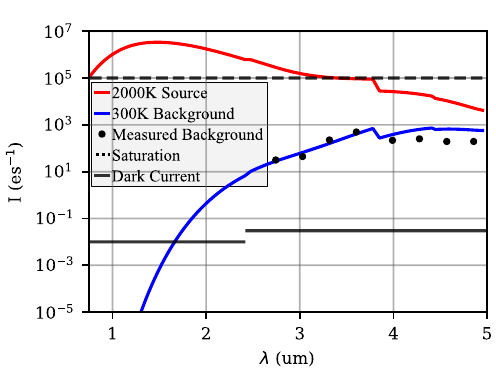}						
        \caption{Detector dark current versus wavelength for the reflected and transmitted channel detector arrays (solid black lines), predicted photocurrent due to the 300 K background from the vacuum window and lab (solid blue line), measured background photocurrent (black dots), predicted photocurrent from a 2000 K blackbody source illuminating a $15 \textrm{ }\mu \textrm{m}$ pinhole array at collimator focus (solid red line), and the photocurrent which results in saturation within a 1 second window (dashed black line). Error bars on the black dots are smaller than the dots themselves and we use neutral density filters on the 2000 K source to avoid saturation at shorter wavelengths.}
        \label{fig:Photocurrent}
    \end{figure}

    Immediately behind the vacuum window we mounted a cold filter, sapphire plate, similar to the vacuum window, but with a $\sim \!\! 100\textrm{ nm}$ thick gold coating on the outside face that reflects infrared radiation (see Figure \ref{fig:ColdFilterAssembly}). For this coating, power reflection/transmission models \cite{FeynmannLecturesVolume2} predict a smooth transmission function of order $10^{-4}$ across the $0.5\textrm{--}5 \; \mu\textrm{m}$ band. Figure \ref{fig:FilterTransmission} shows the measured transmission for the coated cold filter. The cold filter attenuates the background at longer wavelengths to $\sim \!\! 1000$ electrons per second, which allows exposures up to $\sim \!\! 100 \textrm{ s}$ before the detectors saturate (see Figure \ref{fig:Photocurrent}). We typically use $\sim \!\! 10 \textrm{ s}$ exposures for focus measurements.		
    
    The cold filter clamps between two molybdenum rings with roughly the same thermal contraction as sapphire. Thin indium 

    \noindent
    gaskets between the rings and the sapphire take up the small differential contraction when the filter is cooled. The indium also ensures good thermal contact with the sapphire. Titanium flexures support the cold filter by attaching one of the molybdenum rings to a copper flange bolted to the test chamber 60 K radiation shield \cite{OptoMechSystemsDesignYoder}. Both molybdenum rings have flexible copper braid heat straps that connect to the copper flange to cool the filter. A tight-fitting baffle covers all the gaps between the sapphire and the copper flange to ensure that the inner compartment of the test chamber is dark.

    Temperature gradients in the sapphire cause changes in refractive index that can turn the vacuum window and cold filter into weak lenses. The net effect is a $\sim \!\! 3\textrm{ }\mu\textrm{m}$ change in telescope focus for a $1\textrm{ K}$ radial gradient, so the vacuum window has a thin, gold coating on the inside to reduce gradients driven by its radiation into the test chamber. It is more difficult to drive a 

    \noindent
    thermal gradient in the cold filter because the thermal conductivity of sapphire is very high at low temperature \cite{ThermalConductivityofPolycrystallineSolids}.

\begin{figure}[H]
        \centering
        \includegraphics[scale=1]{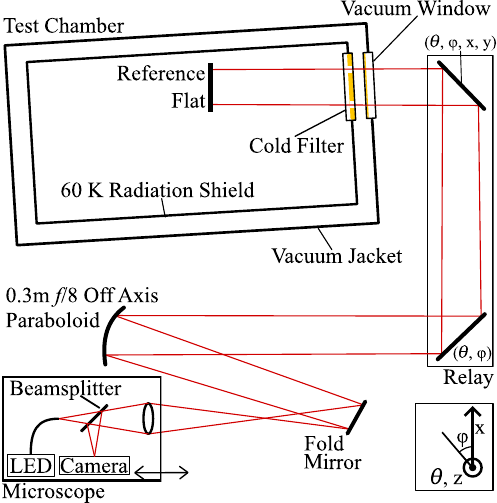}
        \caption{Thermal calibration setup for the cold filter and vacuum window. We mount an uncoated silicon reference flat inside the test chamber and measure focus with the same microscope setup as in Figure \ref{fig:ColFocusSetup}, both before and after cooling the chamber to 60 K. Three uncoated spots on the cold filter form a Hartmann Mask, allowing enough signal to pass through the filter for the microscope camera to detect.}
        \label{fig:WindowRelayCal}
    \end{figure}

    We calibrate the telescope focus shift resulting from thermal loads on the vacuum window and cold filter during focus measurements with a reference flat inside the cold test chamber (see Figure \ref{fig:WindowRelayCal}). Deformation of the flat on cooling can cause systematic changes in measured focus, so we use single crystal silicon to minimize any spatial CTE variations in the reference mirror. We leave the flat uncoated to eliminate stress due to differential contraction of a coating vs. the silicon substrate. Using an interferometer, we measured $<10\textrm{ nm peak-to-valley}$ change in the surface error on cooling the flat inside a small test dewar. The dewar has two shutters: a cold shutter to keep the flat cold and isothermal, and a warm shutter to keep the dewar window warm and isothermal. The shutters are opened just before making a measurement, and we complete the measurement before the flat can deform or thermal gradients can generate optical power in the window.

    In the setup of Figure~\ref{fig:WindowRelayCal}, light passes through the vacuum window and cold filter twice. The optical transmission for this double pass is only $\sim \!\! 10^{-12}$ at longer wavelengths, resulting in very low light levels, well below what the microscope camera can detect. This problem cannot be solved by making the source brighter, because scattering in the microscope limits the dynamic range. Instead, we left three $20\textrm{ mm}$ diameter uncoated spots on the cold filter. In this case, the filter coating becomes a Hartmann mask \cite{Hartmann} with $\sim \!\! 3\%$ optical transmission for a $200\textrm{ mm}$ diameter beam. When the collimator is far from focus, the microscope camera sees three distinct spots with separation that scales 
    
    \noindent
    linearly with focus error. In focus, the camera sees a diffraction pattern with fringe spacing corresponding to the distance between the $20\textrm{ mm}$ spots and overall size corresponding to a $20\textrm{ mm}$ diameter aperture. For telescope focus measurements, the three uncoated spots are covered with opaque tape to prevent the detectors being saturated by $300\textrm{ K}$ radiation through the holes.			

    The Hartmann mask introduces a $dP\sim \!\! 3\times10^{-5}\textrm{ m}^{-1}$ change in optical power because the $20\textrm{ mm}$ diameter beams see power in the window and relay at the positions of the three holes, while the telescope focus measurements see the average error over a $200\textrm{ mm}$ diameter beam. We measured this effect with the reference flat, relay, and an uncoated window assembly in the same configuration as Figure \ref{fig:WindowRelayCal}, but without the chamber. We compared full-aperture measurements to measurements with a Hartmann mask that mimics the openings in the cold filter coating. We apply a $5 \textrm{ } \mu \textrm{m}$ correction to the telescope focus to account for the difference between the Hartmann and full beam shifts.

\begin{figure}[H]
        \centering
        \includegraphics[scale=1]{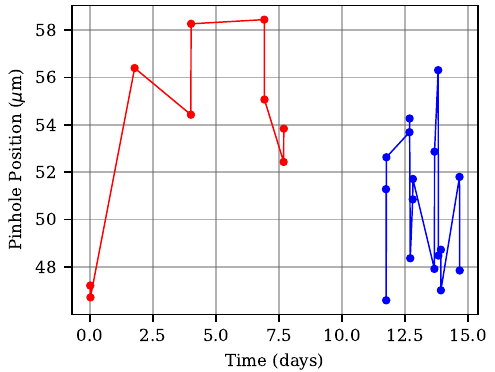}
        \caption{Best focus vs. time with the setup of Figure \ref{fig:WindowRelayCal}, through the Harmann mask. Red points are prior to cooling and blue points are taken with the chamber at $\sim \!\! 60 \textrm{ K}$. The vertical axis corresponds to the displacement of the pinhole array measured in units at telescope focus.}
        \label{fig:ColdFlatFocusVsTime}
    \end{figure}

    Figure \ref{fig:ColdFlatFocusVsTime} shows measured pinhole best focus positions for the vacuum window and cold filter thermal calibration. Measurements were taken over the course of 15 days showing collimator focus to be stable at the $5 \; \mu\textrm{m}$ level. Essentially no change in focus on cooling the chamber was observed, confirming our expectation that the silicon reference flat does not deform and that changes in the optical power of the window assembly are small.

    Optical power in the vacuum window and cold filter also arises from variations in refractive index sourced by a radial dependence of impurities in the sapphire during crystal growth. 

    \begin{figure}[H]
        \centering
        \includegraphics[scale=1]{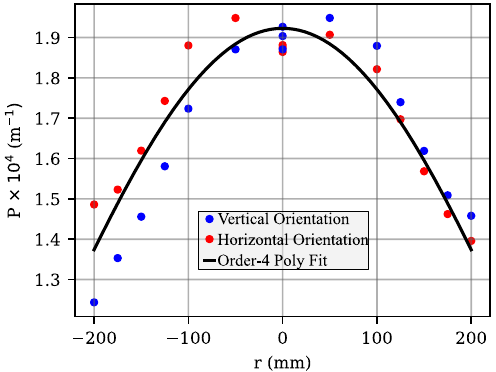}
        \caption{Optical power versus radial position of the 200 mm optical beam from the center of the vacuum window and cold filter assembly. Red/blue dots indicate measurements with the assembly in a horizontal/vertical orientation. The solid black line is an even fourth-order polynomial fit with $\textrm{P} = 1.92\times10^{-4} - 1.56\times10^{-9}r^2 + 4.72\times10^{-15}r^4$, where P is in m$^{-1}$ and $r$ is in mm. Measurements were performed by slipping the window and filter assembly in and out of the beam in the setup of Figure \ref{fig:ColFocusSetup}.}
        \label{fig:WindowPowerVsRadius}
    \end{figure}
    
    \begin{figure}[h]
        \centering
        \includegraphics[scale=1]{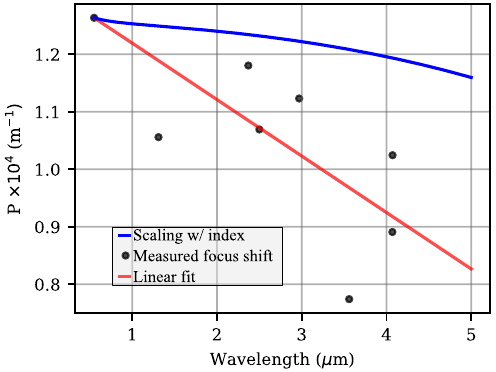}
        \caption{Optical power versus wavelength of the vacuum window and cold filter assembly. Black dots indicate measurements obtained by slipping a third spare sapphire window in and out of the beam during telescope focus measurements (Figure \ref{fig:TelFocusSetup}). The first point at $\lambda = 0.55 \textrm{ } \mu \textrm{m}$ was measured with the Figure \ref{fig:ColFocusSetup} setup. The solid blue line shows the expected optical power upon scaling the measured P$(\lambda = 0.55 \mu \textrm{m})$ value linearly with published values of the sapphire index of refraction \cite{SapphireIndex}. The red line shows a linear fit to the measured points, constrained to match the P$(\lambda = 0.55 \mu\textrm{m})$ value. The fit is P$= 1.23\times10^{-4} - 9.82\times10^{-6}(\lambda - 0.55)$, with P in m$^{-1}$ and $\lambda$ in $\mu$m.}
        \label{fig:WindowPowerVsWavelength}			
    \end{figure}

    \noindent
    We measure this effect by inserting the uncoated window assembly into the collimator beam in the setup of Figure \ref{fig:ColFocusSetup} and translating the assembly to map out the optical power with radius. At the center we observe $P \sim2\times10^{-4} \textrm{ m}^{-1}$ with a $\sim30\%$ decrease in optical power from the center to the edge (see Figure \ref{fig:WindowPowerVsRadius}). The measurement cannot be done after coating because the loss in the cold filter is too high.
    
    We also investigate how the optical power of the window assembly varies with wavelength. The refractive index of sapphire decreases $\sim \!\! 10\%$ across the $\lambda = 0.5 \textrm{--}5.0 \; \mu\textrm{m}$ band \cite{SapphireIndex}, so the optical power of the window assembly should in turn decrease with the wavelength. This effect was measured by inserting a spare, uncoated sapphire window into the collimator beam in the setup of Figure 3 and again during telescope focus measurements. The measured $dP/d\lambda$ is $\sim \!\! 3\times$ larger than expected for a simple scaling with refractive index by $P(\lambda)/P(\lambda=0.55\textrm{ }\mu\textrm{m})$ (see Figure~\ref{fig:WindowPowerVsWavelength}).

    Reflections from the vacuum window and cold filter surfaces generate ghost images. To reduce confusion between the ghosts and the real image, we arranged the vacuum window and cold filter so that no surfaces are parallel to each other, or normal to the collimator beam. The vacuum window and cold filter both have 3--5~arcmin wedges, the cold filter is tilted $1^{\circ}$ relative to the vacuum window, and the combined vacuum window plus cold filter assembly is tilted $8^{\circ}$ relative to the telescope boresight.   
  
\section{Results From Telescope Focus Measurements} \label{sec:Results}	

\begin{figure}[h]
        \includegraphics[scale=1]{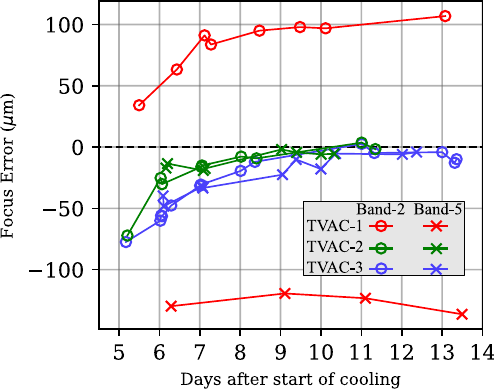}
        \caption{Focus error versus time for all three sets of cold focus measurements (TVACs) on telescope boresight for the reflected (Band-2) and transmitted channel (Band-5) focal planes. The telescope is still cooling through day 9 but achieves a stable temperature thereafter. The focal planes were shimmed after TVAC1 and remained stable after the vibration test between TVAC2 and TVAC3.}
        \label{fig:FocusSettling}
    \end{figure}

\begin{figure*}[h]
        \centering
        \includegraphics[width=\linewidth]{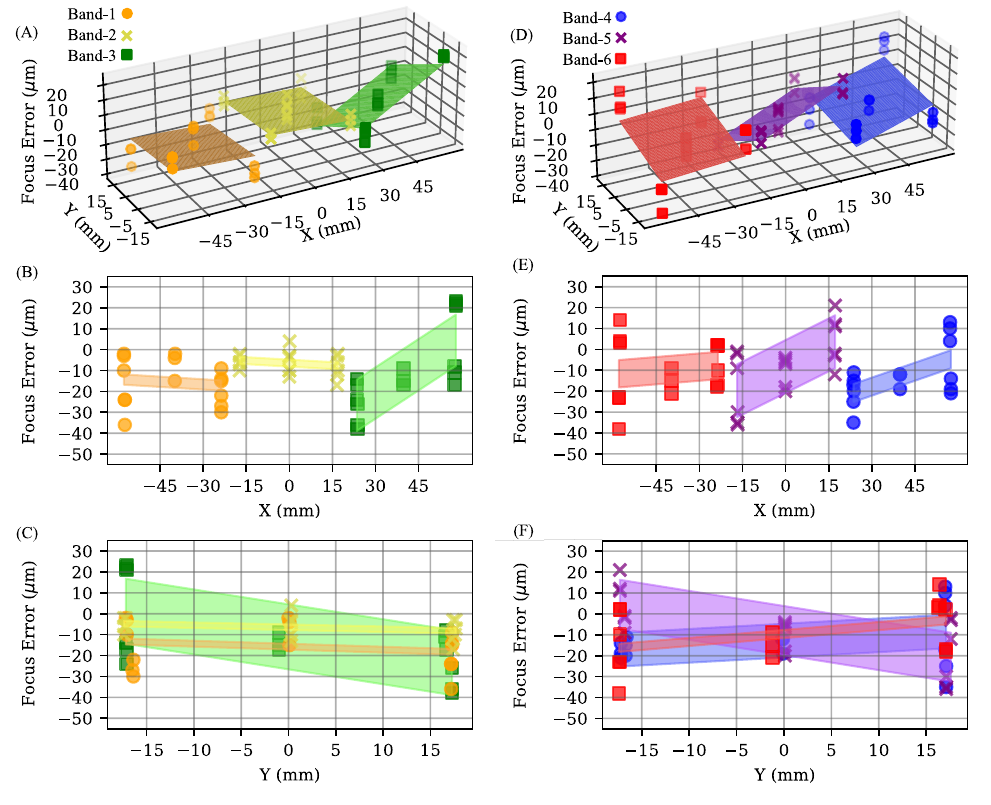}
        \caption{Measured focus errors for all six bands during TVAC-3. Planes are fit to points measured within each individual band to show a tipped detector surface profile. In practice, departures from best focus over the focal surface are a combination of detector piston and tip errors, and optical aberrations in the telescope optics that are most prominent at the corners of the field of view.}
        \label{fig:TVAC3FocusError}
    \end{figure*}

    We performed three sets of cold focus measurements with the SPHEREx telescope. The first (TVAC-1) determined the initial focus error across the field of view. Detector positions were then adjusted using shims in the focal plane. The second measurement (TVAC-2) confirmed that the shims changed focus as expected (see Figure \ref{fig:FocusSettling}). The third and final measurement (TVAC-3) was performed after a random vibration test to simulate the mechanical stress of launch. We saw no change in focus due to vibration. Figure \ref{fig:TVAC3FocusError} shows measured focus errors across the entire field of view for the reflected and transmitted channel focal planes during TVAC-3. We observed stability in the measured focus after $\sim \!\! 9$ days of cooling, upon which repeated measurements were consistent within $\sim5 \textrm{ } \mu\textrm{m}$.

\section{Systematic Error Budget}
    We quantify the accuracy of our focus measurements in terms of systematic focus shifts and our estimates of the error after correction, as summarized in Table \ref{tab:FocusErrorBudget}. The total error of $15.1 \textrm{ } \mu \textrm{m}$ corresponds to $\sim \!\! 1/4$ of the $50 \textrm{ } \mu\textrm{m}$ depth of focus of the telescope. Errors after correction of the window assembly focus shifts are estimated through the RMS of the difference between our models and the measured points shown in Figures \ref{fig:WindowPowerVsRadius} and \ref{fig:WindowPowerVsWavelength}.

\section{Warm Focus Measurements}

\begin{figure}[H]
        \centering
        \includegraphics[scale=1]{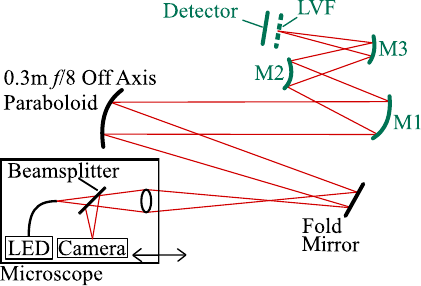}
        \caption{Autostigmatic ``warm'' focus measurement setup to measure position of the linear variable filter front surface relative to best telescope focus.}
        \label{fig:WarmFocusSetup}
    \end{figure}

    As a check on the cold focus measurements described in Section \ref{sec:ColdFocusMeasurements}, we developed a technique to measure the positions of the linear variable filters, that are mounted just in front of the detectors, relative to best focus. We generated a spot on the front surface of a linear variable filter, which is highly reflective at visible wavelengths, using the autostigmatic microscope \cite{AutostigmaticMScope} setup of Figure \ref{fig:WarmFocusSetup}. We inspected the returned image of the spot with the collimator microscope and found the microscope position that minimized the returned spot area. In this measurement, tilts in the reflecting surface have no effect on image position, so the microscope sees a stable, centered spot, with a shift from collimator focus that depends on the displacement of the reflecting surface from telescope focus. Everything in the setup remains at room temperature, we use no vacuum window or cold filter, and there is no need for a relay, because the telescope is mounted on a light stand that provides adjustment of the height, azimuth, and elevation. The simple optical system avoids most of the corrections needed for cold measurements---the only calibration is finding collimator focus with a test flat (see Figure \ref{fig:ColFocusSetup}). The warm focus measurements provide a useful cross check, but they differ from the cold focus measurements at the $100\textrm{ }\mu\textrm{m}$ level at telescope focus, because of fabrication and assembly tolerances in the linear variable filter mounts, cupping of the detector active surfaces, and deformation of the telescope on cooling. This method can also check that shifts in the focal surface position after shimming match expectation.		

    \begin{table}[H]
        \caption{Systematic focus measurement shifts and residual errors after correction, quoted at telescope focus: 1) repeatability in positioning of the fiber and camera in the auto-collimating microscope setup in Figure \ref{fig:ColFocusSetup}; 2) difference between the focus with the reference flat in the cooled test chamber (Figure \ref{fig:WindowRelayCal}) and the sum of power measured in individual components; 3) focus shift due to choice of threshold ($10 \textrm{ to } 25\%$ of the peak) in the spot area calculation for auto-collimating microscope images; 4) scatter in best focus from the reference flat with the chamber cooled (Blue points in Figure \ref{fig:ColdFlatFocusVsTime}); 5) focus shift due to radially-dependent optical power in the vacuum window and filter (Figure \ref{fig:WindowPowerVsRadius}); 6) focus shift due to wavelength-dependent optical power in the vacuum window and filter (Figure \ref{fig:WindowPowerVsWavelength});  7) focus variations from different choices of spot-area calculation and focus curve fitting techniques.}
        \centering
        \begin{tabular}{p{0.4\columnwidth}p{0.2\columnwidth}p{0.2\columnwidth}}
            \hline
            Contribution & Focus Shift ($\mu\textrm{m}$) & Error ($\mu\textrm{m}$)
            \tabularnewline
            \hline
            1) Microscope alignment & 0 & 1.7
            \tabularnewline
            2) Collimator calibration & 0 & 5.6
            \tabularnewline
            3) Collimator best focus calculation & 0 & 4.5
            \tabularnewline
            4) Collimator stability & 0 & 5.6
            \tabularnewline
            5) Power vs. radius for window assembly & 66.9 & 8.9
            \tabularnewline
            6) Power vs. $\lambda$ for window assembly & 16.7 & 5.6
            \tabularnewline
            7) Telescope focus measurement & 0 & 5.6
            \tabularnewline
            TOTAL & 83.6 (sum) & 15.1 (quadrature sum)
            \tabularnewline
            \hline
        \end{tabular}
        \label{tab:FocusErrorBudget}
    \end{table}

\section{Conclusion}

    We have developed a technique for measuring focus errors in a cryogenic, wide-field, near-infrared space telescope. The measurement requires a large, sapphire vacuum window and cold filter to couple light from a collimator into the test chamber that contains the telescope. The vacuum window and cold filter both introduce wavefront errors that result in up to $\sim \! 83.6 \textrm{ }\mu\textrm{m}$ of telescope focus shift, which we account for and achieve a residual systematic error after correction of $\sim \! 15 \textrm{ } \mu \textrm{m}$. We performed measurements on the SPHEREx telescope, and achieved results with a statistical accuracy of $\sim 5 \textrm{ } \mu \textrm{m}$. The accuracy of our focus measurements is predominantly limited by the precision of our characterization of wavefront error in the vacuum window and cold filter, which in turn is limited by the mechanical stability of our setup and seeing in the lab environment. Potential improvements include the fabrication of windows with a smaller transmitted wavefront error and reducing the overall length of the system such that the relay mounts on the same optical bench as the source and paraboloid.

\begin{backmatter}
    \bmsection{Funding}	
    National Aeronautics and Space Administration (80NM0018D0004).

    \bmsection{Acknowledgements}
    We thank Darren Dowell, Kenneth Mannat, Hien Nguyen, and Marco Viero for productive discussions regarding the collimator design and for helpful comments on the manuscript. The research described in this paper was carried out at the California Institute of Technology under a contract with the National Aeronautics and Space Administration.
        
    \bmsection{Disclosures}
    The authors declare no conflicts of interest.
    \bmsection{Data Availability}
    Data underlying the results presented in this paper are not publicly available at this time but may be obtained from the authors upon reasonable request.
\end{backmatter}

\bibliography{references}

\end{document}